\documentclass{sigchi}

\usepackage{balance}  
\usepackage{graphics} 
\usepackage{times}    
\usepackage{url}      

\makeatletter
\def\url@leostyle{%
  \@ifundefined{selectfont}{\def\UrlFont{\sf}}{\def\UrlFont{\small\bf\ttfamily}}}
\makeatother
\def\pprw{8.5in}
\def\pprh{11in}

\usepackage[pdftex]{hyperref}
\hypersetup{
pdftitle={SIGCHI Conference Proceedings Format},
pdfauthor={LaTeX},
pdfkeywords={SIGCHI, proceedings, archival format},
bookmarksnumbered,
pdfstartview={FitH},
colorlinks,
citecolor=black,
filecolor=black,
linkcolor=black,
urlcolor=black,
breaklinks=true,
}

\begin{document}

\title{CrowdCafe - Mobile Crowdsourcing Platform}

\numberofauthors{6}

\author{
  \alignauthor Pavel Kucherbaev\\
    \affaddr{University of Trento}\\
    \affaddr{Via Sommarive 9, Povo (TN)}\\
    \affaddr{38123, Italy}\\
    \email{pavel.kucherbaev@unitn.it}
  \alignauthor Azad Abad\\
    \affaddr{University of Trento}\\
    \affaddr{Via Sommarive 9, Povo (TN)}\\
    \affaddr{38123, Italy}\\
    \email{azad.abad@unitn.it}
  \alignauthor Stefano Tranquillini\\
    \affaddr{University of Trento}\\
    \affaddr{Via Sommarive 9, Povo (TN)}\\
    \affaddr{38123, Italy}\\
    \email{stefano.tranquillini@unitn.it}
  \alignauthor Florian Daniel\\
    \affaddr{University of Trento}\\
    \affaddr{Via Sommarive 9, Povo (TN)}\\
    \affaddr{38123, Italy}\\
    \email{florian.daniel@unitn.it}  
  \alignauthor Maurizio Marchese\\
    \affaddr{University of Trento}\\
    \affaddr{Via Sommarive 9, Povo (TN)}\\
    \affaddr{38123, Italy}\\
    \email{maurizio.marchese@unitn.it}  
  \alignauthor Fabio Casati\\
    \affaddr{University of Trento}\\
    \affaddr{Via Sommarive 9, Povo (TN)}\\
     \affaddr{38123, Italy}\\
    \email{fabio.casati@unitn.it}
}

\maketitle

\begin{abstract}

In this paper we present a mobile crowdsourcing platform CrowdCafe, where people can perform microtasks using their smartphones while they ride a bus, travel by train, stand in a queue or wait for an appointment. These microtasks are executed in exchange for rewards provided by local stores, such as coffee, desserts and bus tickets. We present the concept, the implementation and the evaluation by conducting a study with 52 participants, having 1108 tasks completed.

\end{abstract}



\section{Introduction}

Crowdsourcing is the practice of outsourcing work to an unknown group of people via the Internet, instead of assigning it to internal employees \cite{Howe2006}. Crowdsourcing has been so far very successful in performing tasks which are still hard to automate using algorithms, while they can be relatively easily solved by humans, such as image object recognition, annotations, feedback collection and similar.

Requestors are the people who want to crowdsource their work. They publish tasks on crowdsourcing platforms where requestors meet potential workers - people who solve tasks for monetary reward, curiosity or other motivations. Some examples of crowdsourcing platforms are Amazon Mechanical Turk (MTurk), CrowdCloud, MicroWorkers, Mobileworks, CrowdFlower. In general workers need to perform tasks from their desktops or laptops, as tasks are designed for non-mobile screens. However, many people, when they have time with their computer, prefer to perform some things related to their job or just to have fun watching something. 

People spend everyday some amount of time riding a bus, standing in a line at a grocery store, waiting for a doctor appointment. During this time they can read a pocket book or use their smartphones. Many people end up checking their social network profiles. Mea et al. \cite{Mea2013} showed an evidence that users can perform some tasks via mobile devices faster than via desktop. 


We present a crowdsourcing platform CrowdCafe, where people perform microtasks, specifically designed for mobile execution, in exchange for non-monetary rewards provided in local stores, such as coffee, desserts and bus tickets. 



\section{State of the art}
The current research in the field of mobile crowdsourcing can be separated by three objectives: i) to help people from developing countries to earn extra cash, ii) to utilize smartphone sensors to collect location specific data and iii) to discover new concepts of performing crowdsourcing tasks.

\paragraph{Helping people from developing countries}
Eagle et al. \cite{Eagle2009} and Kulkarni et al. \cite{Kulkarni2012} presented platforms (\emph{txteagle} and \emph{MobileWorks}), using which people from developing countries can earn extra money by completing various tasks using their mobile low-cost phones. Gupta et al. \cite{Gupta2012} presented a platform \emph{mClerk} for mobile paid crowdsourcing in developing regions, which processes (sends and receives) tasks via SMS.

\paragraph{Mobile-sensing}
Yan et al. \cite{Yan2009} proposed an iPhone-based mobile crowdsourcing platform \emph{mCrowd}, using which mobile users can perform tasks, using their smartphone sensors.
Tamilin et al. \cite{Tamilin2012} presented a context-aware crowdsourcing system for conducting crowdsourcing campaigns with smart phone users, which utilizes sensors available on mobile devices.


\paragraph{Discovering new concepts}
Vaish et al. \cite{vaishtwitch} presented an Android application \emph{Twitch}\footnote{\url{http://twitch.stanford.edu/}}, which in order to unlock a phone, asks its owner to answer a simple question, such as: how many people are around, or which activity the owner is doing now. Similarly, Truong et al. \cite{Truong2014} showed how different crowdsourcing tasks can be completed using different unlocking gestures.
Heimerl et al. \cite{Heimerl2012} presented \emph{Umati} -- communitysourcing vending machine, which helps to attract a specific local group of people (e.g. people with deep knowledge in computer science) to perform tasks on the screen of the vending machine in exchange for snacks.
Luon et al. \cite{Luon2012} proposed a mobile system \emph{Rankr} for crowdsourcing opinions via pair comparison of images and sentences on mobile phones. 
In \cite{Kittur2013} Kittur et al. analyzed how different aspects of crowdsourcing could be improved, they also challenged the community to revolutionize the conception of what a crowdsourcing platform is.

Musthag et al. \cite{Musthag2013} did an analysis of differences between mobile crowdsourcing platforms and desktop ones. They found a significant difference in demographics. The comparison was not straight in sense that mobile crowdsourcing platforms mostly support offline location-dependent tasks, while desktop support online mainly.
Mea et al. \cite{Mea2013} conducted user studies where they tried to identify which crowdsourcing tasks suit better for mobile and which for desktop devices.


With CrowdCafe we aim: 
\begin{itemize}
\item to investigate how during semi-occupied situations (such as riding a bus, traveling by train or waiting in a line) people can perform microtasks using their smartphones, not for the purpose of making income, but to have fun and to benefit out of this time,
\item to boost the research in the mobile crowdsourcing field, by providing to the academic community an open-sourced platform which is deployed online, so other researchers can conduct studies and extend the platform if needed,
\item to identify the best practices of designing tasks user interfaces and to create a repository of reusable user interface patterns.
\end{itemize}



\section{Concept}

The concept of CrowdCafe affects three aspects of crowdsourcing: \emph{tasks user interfaces}, \emph{tasks classification} and \emph{workers motivation}. 

\paragraph{Tasks User Interfaces}

In order to provide a good user experience of tasks execution on CrowdCafe we want to apply the best user interface (UI) practices from current mobile applications, such as: \emph{feed} to present all the content as a list, without sidebars; \emph{big full width buttons} to make it comfortable to press them with a thumb; \emph{swipe for action} to keep a user interface very clean without buttons, where, depending on a direction and a distance of swiping a UI element, different actions are triggered (was announced with the MailBox mobile application\footnote{\url{http://www.mailboxapp.com/}}).

\paragraph{Tasks Classification}
\label{tasks_classification}

As described by \cite{Kucherbaev2014} on the platforms such as MTurk or CrowdFlower it is hard for workers to select a task to work on, because descriptions are not informative enough and they never know how much time they will spend on execution. On CrowdCafe we decided to split all the tasks by completion time in 3 clear categories: 

\begin{itemize}
\item \emph{``Espresso''}  - about 10 seconds to be completed, with mostly only clicking and swiping actions required (e.g. to identify the sentiment of tweets or to compare pairs of images);
\item \emph{``Cappuccino''} - about 2 minutes to be completed, with some typing and learning required (e.g. to fill up a short survey, to annotate images);
\item \emph{``Wine''}  - more than 5 minutes to be completed, might require a worker to be in a specific context or a location (e.g. to go to a grocery store and to make a photo of a particular product with its price). 
\end{itemize}
  
\paragraph{Workers Motivation}

We do not position CrowdCafe as a source for primary or secondary income. We consider CrowdCafe as a way to convert time, which is wasted otherwise, to enjoyable rewards. According to Dan Ariely \cite{Ariely2002} the smaller reward now is more desirable than a bigger one later. So we want to minimize the time frame between a worker starting task execution and a worker enjoying a reward, by providing micro rewards from local stores, such as coffee or dessert at a university bar. This can help workers to start working on tasks and in 15 minutes to feel an outcome of their work by drinking a coffee.
\section{Implementation}

We have implemented the CrowdCafe as a website\footnote{\url{http://crowdcafe.io/}}. It has two main components: i) ``Kitchen'', where requestors design and publish tasks from a desktop; ii) ``Cafe'', where workers perform tasks using their mobile phones and get reward coupons. The CrowdCafe code is open-sourced and is available on GitHub\footnote{\url{https://github.com/CrowdCafe/crowdcafe}}, where more details about the implementation can be found. 

\subsection{Requestors Interface}

\begin{figure}[ht]
    \centering
    \includegraphics[width=0.48\textwidth]{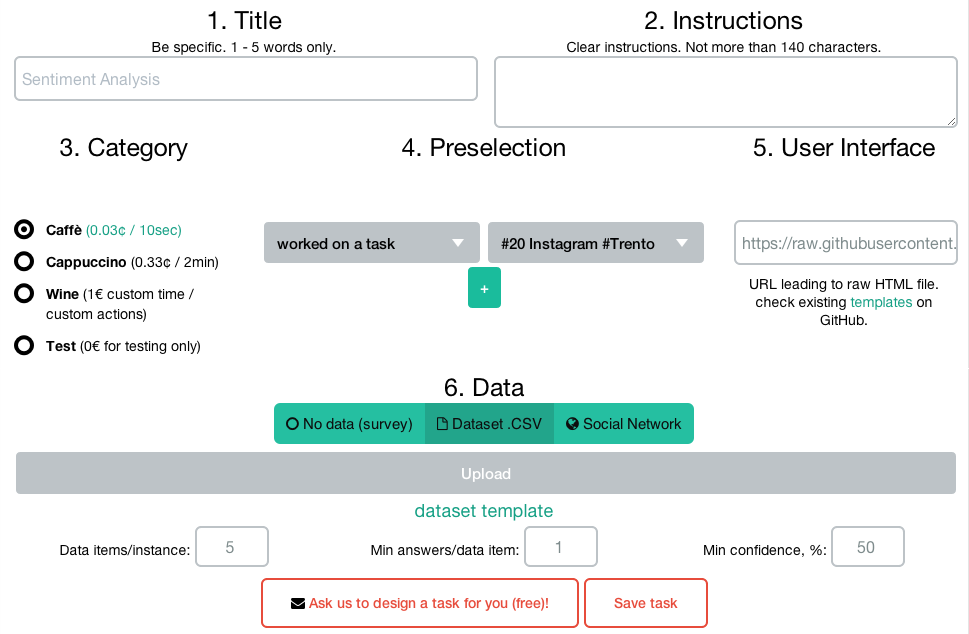}
    \caption{Requestors user interface}
    \label{kitchen_ui}
\end{figure}

In ``Kitchen'' requestors create tasks (Figure \ref{kitchen_ui}), defining: 1) title, 2) instructions, 3) category, 4) preselection logic, 5) user interface, 6) input dataset and 7) quality control settings. The first three parts are trivial and we focus on the other four.
\begin{figure*}[ht]
    \centering
    \includegraphics[width=\textwidth]{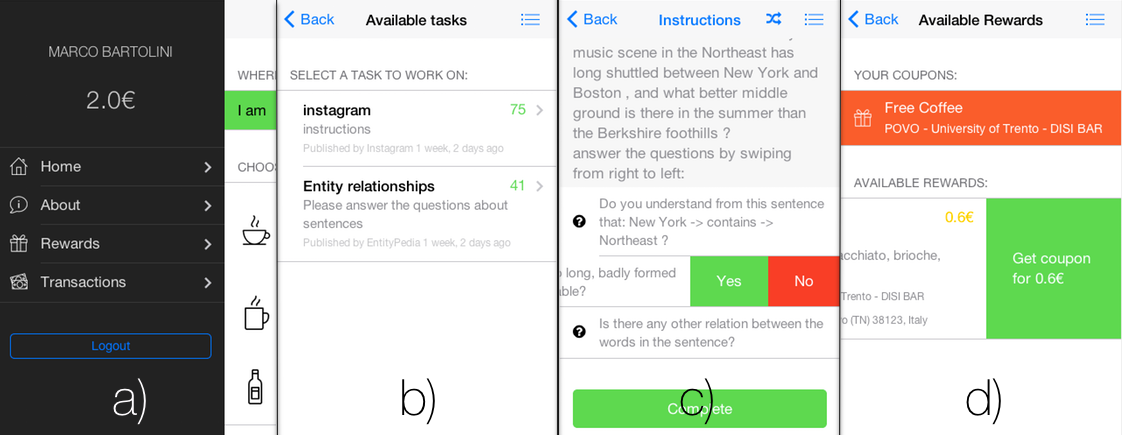}
    \caption{Workers User Interface. a) Task categories and side menu, b) Available tasks, c) Example of a task, d) Reward page}
    \label{cafe_ui}
\end{figure*}
\paragraph{Preselection}
Requestors can define to which workers the task will be visible by adding a set of restrictions, such as ``worked'' or ``did not work'' on particular tasks. This simple preselection logic is powerful enough to route surveys to particular workers, to create tasks which are only visible to workers who performed some skill test tasks.

\paragraph{User Interface}
We decided to leave a lot of freedom for requestors, so they can design the UI of their tasks using HTML and CSS. In order to not design from scratch and to accumulate the best practices such UI templates are stored in the public GIT repository. To apply a particular template, requestors need to insert its URL, which refers to a raw HTML file.

\paragraph{Input Dataset}
There are three options for input data: 1) no input data (a survey task), 2) data uploaded from a .csv file, 3) data from a social network feed (e.g. Twitter, Instagram) on a particular topic, defined by a hash-tag (such as \#helsinki).
Requestors also define how many data units (rows in case of .csv file, or tweets in case of Twitter) a worker should process in one time. 

\paragraph{Quality Control}
Requestors can define a similarity function to check whether a given judgement is similar to gold data (predefined correct answer) or whether judgements given by different workers are similar to each other (an agreement is found). By default this similarity function is a simple equality, while requestor can create a script which does more complex similarity assessment.

Gold units (if available) are injected into tasks with a probability calculated by the formula:
\begin{equation}\label{eq:fourierrow}
p = \frac{1+N_{incorrect}}{1+N_{incorrect}+N_{correct}}
\end{equation}
, where \begin{math}N_{incorrect}\end{math} -- number of incorrect judgements for gold units given by a worker, \begin{math}N_{correct}\end{math} -- number of correct judgements for gold units given by a worker. The more correct judgements a worker gives, the less probability of having them further injected he has.

Requestor can also define a limit of mistakes a worker is allowed to make (which is zero by default).
\subsection{Workers Interface}

The mobile website for tasks execution has four main sections: 1) home page and side menu (Figure \ref{cafe_ui}a), 2) tasks listing page (Figure \ref{cafe_ui}b), 3) task execution page (Figure \ref{cafe_ui}c), 4) reward page (Figure \ref{cafe_ui}d).

\paragraph{Home page}
On the home page workers see the list of task categories. In the right part of the top bar on every page there is a button which opens the side menu. There is a context select box where we ask workers to define where they currently are (e.g. in a bus, having lunch).
On the side menu workers see their name, the amount of money they have earned and four links: 1) ``Home'', 2) ``About'', where the service is described in details, 3) ``Rewards'', 4) ``Transactions'' - where the history of all earnings (related to tasks execution) and spendings (related to purchasing coupons) is presented. This side menu also has a logout button.

\paragraph{Tasks listing page}
On the tasks listing page workers see a list of tasks with its description and amount of instances available.

\paragraph{Task execution page}
When workers start executing a task, they first see a pop up window with instructions. Workers are expected to read them and after they can start to perform a task. Workers should fill up all the necessary fields otherwise they can not submit the task. After they submit one task, they are redirected to another instance of this task. If there are no any instances available workers are redirected back to the tasks listing page.

\paragraph{Rewards page}
On this page a worker can see a list of available rewards with its price and address where they can get them. 

\section{Evaluation}
In order to evaluate the concept of CrowdCafe along with our implementation we decided to post two tasks and one follow up survey:

\begin{itemize}
\item ``Instagram \#Trento'' -- in this task workers were expected to look on images potentially about Trento and provide two actions: 1) add three relevant tags to each image, 2) specify whether this image really represents Trento as a city.

\item ``Sentence Analysis'' -- in this task workers were expected to read a short text and to answer two questions: 1) is a given relationship between two nouns correct from the content of the text? (yes, no, i don't know) 2) does the text consist of only one clear sentence? (yes, no)

\item Survey -- in this task we wanted to collect the workers feedback about tasks and the CrowdCafe platform in general.
\end{itemize} 

For the first task we uploaded a dataset of 1000 sentences, splitting them in 334 tasks of 3 sentences, asking for at least 3 judgments for each sentence. For the second task we took a feed of 231 images from Instagram with a hashtag \#trento, splitting them in 77 tasks of 3 images, asking for at least 3 judgments. Both tasks were qualified as ``Espresso'' tasks with a reward of 0.03 euro. The final survey was qualified as ``Cappuccino'' task with a reward of 0.33 euro.

We prepaid 84 coffees (0.60 euro each) at the bar on our faculty and left there a list of 84 unique codes. When workers earn enough money, they can purchase a coupon which they exchange for a coffee or a dessert at the bar.

In order to approach the first users we sent email invitations to people from our research group (30 people) and distributed 20 printed posters around our faculty building. It helped to get 80 sign ups on the platform in 1 day.


\subsection{Results}

We collected all the judgments for all tasks in two days from 52 workers. Two workers were identified which used the vulnerability in the code and submitted extra 400 equal judgments. These judgments were removed from the analysis. Some people did not specify a place where they performed tasks, so about a half (46.9\%) of all responses did not have associated place. Out of those, which had: 56.40\% were performed on a workplace, 14.10\% -- outside, 13.13\% -- in a bus, 11.83\% -- at home, 4.38\% -- in a train, 0.16\% -- walking.  

\paragraph{Task 1 - ``Instagram \#Trento''}
The average execution time for this task was 107.31 seconds (317 task responses, median 87, standard deviation 88.03 sec). We received 791 judgments for 231 image. There were 737 (93.17\%) images according to instructions and included three or more tags, while 54 included only 1 or 2 tags. 

Kappa evaluation metric shown to be accepted option to analyze the reliability of the inter agreement among workers\cite{carletta1996assessing}. We used Fleiss Kappa \cite{fleiss1971measuring} to assess the reliability of the provided tags. We could not use the tags directly to estimate the Kappa values due to the open vocabulary of the tags provided by workers. Therefore, 3 experts (members of our research group) categorized all the tags into 10 predefined clusters and voting system has been used to select the ground truth cluster for each tag. Finally, cluster names were replaced with the real tags and Kappa values were calculated.
The overall Fleiss Kappa value is 0.4416 with 0.0154 error that is an indication of a moderate agreement. The 95\% confidence interval of Fleiss Kappa is [43.4, 44.9]. Also, the \emph{p-value} is less than 0.0001 which shows that the observed agreement is statistically significant.

\paragraph{Task 2 - ``Sentence Analysis''}
The average execution time for this task was 62.85 seconds (1006 task responses, median 16 sec,  standard deviation 276.76 sec). For each sentence we received from 3 to 5 judgments. These judgments have very low agreement level (we did not find any agreement in the majority of sentences).


\paragraph{Survey task}
We sent email invitations to 18 people who provided at least one judgment to both ``Instagram \#Trento'' and ``Sentence Analysis'' tasks. There are 15 people completed this survey. Out of these people 66\% responded that 0.03 euro is enough reward for such tasks, 80\% responded that they will use the platform in future, the average interest on scale from -3 (very negative) to +3 (very positive) in ``Instagram \#Trento'' is 0.93, in ``Sentence Analysis'' is -0.60. The average overall satisfaction about CrowdCafe platform is 1.93 on the same scale.

\subsection{Discussion}
When we designed our two tasks we did not expect that it would take people so much time (107.31 and 62.85 seconds) to complete them. It showed that we classified these tasks not correct. Still we had workers, providing many judgments and in the survey the majority of workers responded, that 0.03 euro is enough reward for completing such tasks (which is only about 2 euro per hour). Several people mentioned that they performed tasks thinking about coffee and not money.

The overall quality of tags provided in ``Instagram \#Trento'' task is high. All the tags were relevant. Even when workers did not follow the instructions and provided less than 3 tags or tags in other language than English (54 judgments), tags were still relevant. We found very low agreement between workers stating that an image characterizes Trento. There are two possible reasons: 1) instructions were not clear enough and some people marked only images which have some famous Trento building on it, while others marked all images which they believed were made in Trento, 2) some users pointed an issue that this button did not work well in the native Android browser.

In the ``Sentence Analysis'' task there is very low agreement between workers, because many workers did not understand the instructions clearly and some workers did not pay enough attention and simply provided random results. This is also the reason of the big standard deviation and big difference between mean and median execution time for this task. In addition the survey results show that the interest in this task was much lower than in the ``Instagram \#Trento'' task.

\section{Conclusion}
In this paper we have described the concept and evaluated the implementation of the mobile crowdsourcing platform CrowdCafe, where people can and are willing to perform microtasks during short spans of free time in exchange for tangible rewards such as coffee. We showed that for well-designed tasks with clear instructions, even without any specific control (e.g. gold data, skill tests), workers provide results of a very high quality (93.17\%). In tasks with ambiguous instructions the quality of results is poor. 

In future we plan to investigate: i) the variety of tasks people can perform on their smartphones with better or the same quality as on regular ``desktop'' crowdsourcing platforms (as an extension of Mea et al. \cite{Mea2013} work), ii) the workers productivity with different user interface approaches in tasks design (e.g. radio buttons, swiping to action, set of buttons), iii) how different motivation strategies (e.g. cash, coffee, donation to charity, no reward) affect the workers productivity.


%
%
%
%
%
\balance
\bibliographystyle{acm-sigchi}
\bibliography{bibliography}
\end{document}